# The Visual Monitoring Camera (VMC) on Mars Express: a new science instrument made from an old webcam orbiting Mars

**Jorge Hernández Bernal**[1,6], **Alejandro Cardesín Moinelo**[2,7,8], **Ricardo Hueso**[1], **Eleni Ravanis**[2,9], **Abel Burgos Sierra**[2], **Simon Wood**[3], **Marc Costa Sitja**[2], **Alfredo Escalante**[2], **Emmanuel Grotheer**[2], **Julia Marín Yaseli de la Parra**[2], **Donald Merrit**[2], **Miguel Almeida**[4], **Michel Breitfellner**[2], **Mar Sierra**[2], **Patrick Martin**[2], **Dmitri Titov**[5], **Colin Wilson**[5], **Ethan Larsen**[1], **Teresa del Río Gaztelurrutia**[1], **Agustín Sánchez Lavega**[1]

1. Dpto. Física Aplicada, Escuela de Ingeniería de Bilbao, Universidad del País Vasco UPV/EHU, Bilbao, Spain
2. ESAC, European Space Agency, Madrid, Spain
3. ESOC, European Space Agency, Darmstadt, Germany
4. Dias Almeida Data Processing and Systems, DADPS, Bern, Switzerland
5. ESTEC, European Space Agency, Noordwijk, Netherlands
6. Laboratoire de Méteorologie Dynamique, Sorbonne Université, Paris, France
7. Instituto de Astrofisica e Ciencias do Espaço, Universidade de Lisboa, Lisbon, Portugal
8. Instituto de Astrofísica de Andalucía, IAA-CSIC, Granada, Spain
9. Hawaiʻi Institute of Geophysics and Planetology, University Of Hawaiʻi at Mānoa

**Corresponding author: Jorge Hernández-Bernal, jorge.hernandez@ehu.eus**

## Highlights

- The preparation of an engineering camera to be used as a science instrument is described.
- Developed procedures enabled this new fully operational science instrument.
- Low-cost cameras can expand the science return of current and future space missions.

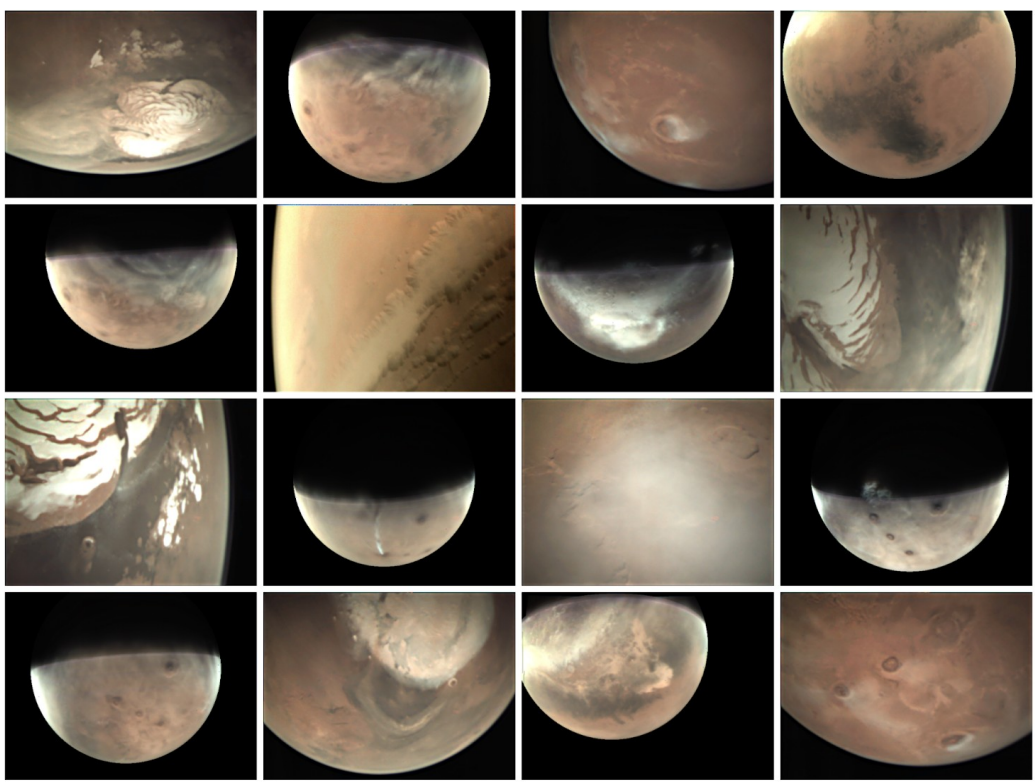

## Abstract

The Visual Monitoring Camera (VMC) is a small imaging instrument onboard Mars Express with a field of view of ~40º×30º. The camera was initially intended to provide visual confirmation of the separation of the Beagle 2 lander and has similar technical specifications to a typical webcam of the 2000s. In 2007, a few years after the end of its original mission, VMC was turned on again to obtain full-disk images of Mars to be used for outreach purposes. As VMC obtained more images, the scientific potential of the camera became evident, and in 2018 the camera was given an upgraded status of a new scientific instrument, with science goals in the field of Martian atmosphere meteorology. The wide Field of View of the camera combined with the orbit of Mars Express enable the acquisition of full-disk images of the planet showing different local times, which for a long time has been rare among orbital missions around Mars. The small data volume of images also allows videos that show the atmospheric dynamics of dust and cloud systems to be obtained. This paper is intended to be the new reference paper for VMC as a scientific instrument, and thus provides an overview of the updated procedures to plan, command and execute science observations of the Martian atmosphere. These observations produce valuable science data that is calibrated and distributed to the community for scientific use.

## 1. Introduction

Mars Express (MEX; Chicarro et al., 2004) arrived at Mars in 2003 and became the first interplanetary probe of the European Space Agency (ESA). Its initial scientific payload included a set of seven instruments, including two imagers (HRSC, Neukum et al., 2004; OMEGA, Bibring et al., 2004), a radar (MARSIS, Picardi et al., 2004), two spectrometers for atmospheric studies (PFS, Formisano et al., 2004; SPICAM, Bertaux et al., 2004), a particle detector (ASPERA, Barabash et al., 2004), and a radioscience experiment (MARS, Pätzold et al., 2004). At the time of the design of MEX, ESA was introducing small Visual Monitoring Cameras (VMCs) on its spacecraft for engineering and outreach purposes (Habinc et al., 2000), and one of them (VMC from now onward) was installed on MEX (Ormston et al. 2008) with the purpose of visually monitoring the separation of the Beagle 2 lander (Pullan et al., 2004).

Years after the initial mission of VMC to monitor the separation of Beagle 2, in 2007, it was recommissioned to be used as an outreach camera that would be able to obtain full-disk images of Mars attractive to the general public (Ormston et al., 2008; 2011). The observations obtained in that period were acquired in a fully opportunistic way with no scientific intent, but they showed the potential of VMC to complement the payload on Mars Express (Schulster et al., 2011). In 2016 VMC started to be exploited as a science instrument, demonstrating its scientific potential to observe atmospheric phenomena (Sánchez-Lavega et al., 2018a). In May 2018 its science operations started to be coordinated by the MEX Science Ground Segment (Merrit et al., 2018) in close collaboration with the science team (Ravanis et al., 2019). From 2018 onwards, the number of observations acquired by VMC increased significantly, producing tens of thousands of images, and allowing a wide range of new atmospheric studies (Sánchez-Lavega et al., 2024). Topics addressed by VMC investigations were the study of the dynamics of dust storms (Hernández-Bernal et al., 2019; Sánchez-Lavega et al., 2022a), the characterisation of clouds (Sánchez-Lavega et al., 2018b; Hernández-Bernal et al., 2021a; 2021b), the monitoring of the evolution of polar ice caps (Río-Gaztelurrutia et al., 2021), the support to in-situ measurements from surface landers and rovers (Sánchez-Lavega et al., 2022b), and the potential to search for meteors (Hernández-Bernal et al., 2021c).

VMC has thus become the eighth science instrument on Mars Express, and is now managed in a way similar to that of the other instruments in the spacecraft. Images are processed, calibrated and have been regularly archived in the ESA Planetary Science Archive (PSA; Besse et al., 2018) since September 2020 for distribution to the science community (Ravanis et al., 2020). Higher level science products are produced for specific sets of observations through a combination of near consecutive images with different exposures following an image fusion procedure (Hernández-Bernal et al., 2019; 2021a).

The basic technical description of VMC, its early recommissioning history, and the outreach experiences, were already described by Ormston et al. (2008; 2011), and are also documented in the VMC blog[1]. The main purpose of this paper is to serve as a new reference for the scientific operations of VMC and the technical aspects required in data analysis of its images, similarly to published references for other MEX instruments (e.g. Scholten et al., 2005; Belucci et al., 2006). The main challenges in transforming what was initially an engineering instrument into a full science instrument are described here. The main challenges were related to the photometric and geometric calibration of VMC images without ground-based calibration data. The current status of the instrument allows regular deliveries of the data to ESA's PSA (Cardesín-Moinelo et al., 2024).

This paper is organized as follows. In section 2 we introduce VMC, describing its limitations and capabilities, and we summarize the science results obtained from VMC images so far. In section 3 we describe the VMC operations and how they have been optimized to increase its scientific capabilities. In section 4 we describe VMC science data, including the photometric calibration and the main artifacts that are still present in the images. We finish this section by describing a method to produce High Dynamic Range (HDR) images from VMC data. In section 5 we describe the calibration of geometry, which mainly involves determining the accurate orientation of VMC relative to the spacecraft. In section 6 we give some closing remarks and conclusions.

## 2. A new science instrument onboard Mars Express

### *2.1. VMC instrument description*

The VMC hardware is described in the VMC Flight User Manual (Jameux, 2003; Flight User Manual from now onwards). Here we provide the key reference data extracted from that document and we update some characteristics of the hardware with the most accurate knowledge based on the calibration process and the experience gained while using the instrument.

Fig. 1 shows ground photographs of VMC during assembly, and schematics of the instrument as extracted from the Flight User Manual. The main characteristics of VMC are summarized in Table 1.

---

[1] https://blogs.esa.int/vmc/blog/

The VMC sensor is a CMOS with 640x480 pixels covered with a Bayer mask filter to allow for color, which was a typical sensor for a webcam in the 2000s. Considering the Field of View (FOV) of the instrument and the MEX orbit around Mars, this translates to a nadir resolution of ~11 km/pixel from apocenter, and ~0.4 km/pixel from pericenter (though observations from pericenter are less common). The color depth of the sensor is only 8 bits, which is low compared to most scientific cameras on planetary missions; in sections 3.2 and 4.3 we describe how we optimized operations and data processing to partially overcome this limit and increase the dynamic range of the observations.

Given the way VMC is connected to the onboard computer, all other science instruments must be turned off when VMC is observing. This is a major constraint that limits the VMC operations. In addition, VMC is tilted 19º (around X axis) from the Z axis of MEX, which is the direction in which all the other instruments are pointed. As a result, VMC observations require a specific pointing of MEX, with an offset with respect to the standard nadir boresight. On the other hand, the low data volume of VMC images implies a minimal memory allocation and downlink time (see table 1) compared to other instruments.

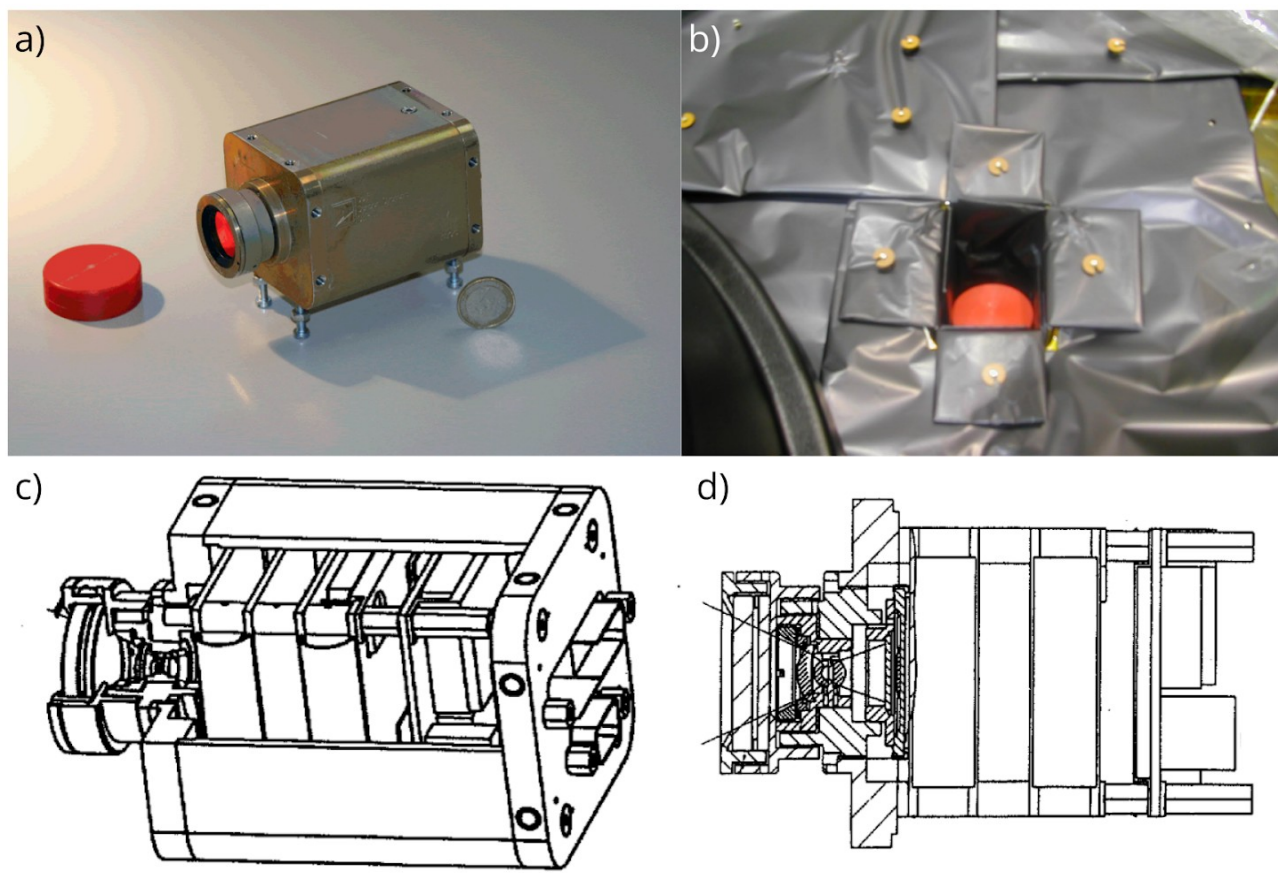

Fig. 1. Ground photos and schemes of VMC. a) Photo of the instrument before launch, a coin was included to illustrate the size of the instrument, the red "remove before flight" cover is in the left side of the image. b) Ground photo of VMC mounted on MEX, only the red "remove before flight" cover is visible. c and d) Schemes of VMC as reproduced from the Flight User Manual.

Another limitation of VMC is the low speed at which data is able to be transferred to the onboard computer of MEX. Each image takes 38 seconds to be transferred, and no other scientific operation can be done in parallel. Additionally, a margin of a few seconds is left to prevent superposition of different telecommands. As a result, the time between individual images is typically around 48 seconds.

VMC only sends a flux of bits containing the image to the onboard mass memory, without additional housekeeping data from the instrument. Thus, the exact timestamp of images is unknown. In general, it has been assumed that the time when the first packet of an image was received by the onboard computer could be used as the image timestamp. However, we have found that this timing of images is not accurate, as we discuss in section 5.2.

Table 1. Main VMC characteristics. Values extracted from the Flight User Manual, with the exception of the FOV, for which the calibrated value is given (see section 5.1)

| Number of pixels | 640 x 480 |
|---|---|
| Color depth | 8 bits |
| Raw image size | 307 200 bytes |
| Spectral capabilities | RGB Bayer Pattern |
| Pixel size | 14 µm |
| Focal length | 12.3 mm |
| Field of View (FOV)* | 39.75º x 30.34º |
| Transfer to memory time | 38 s |
| Power consumption | <3W (up to 5W during capture) |
| Exposure times | 0.4 ms – 90 s |
| Weight | 430 g |
| Instrument size | 65x60x108 mm |

## 2.2. Scientific targets. What is possible with a "webcam" orbiting Mars?

VMC clearly does not have the same technical capabilities of other instruments on planetary missions, however, the configuration of VMC on MEX enables some science observations that in some cases are not possible using other instruments, described here.

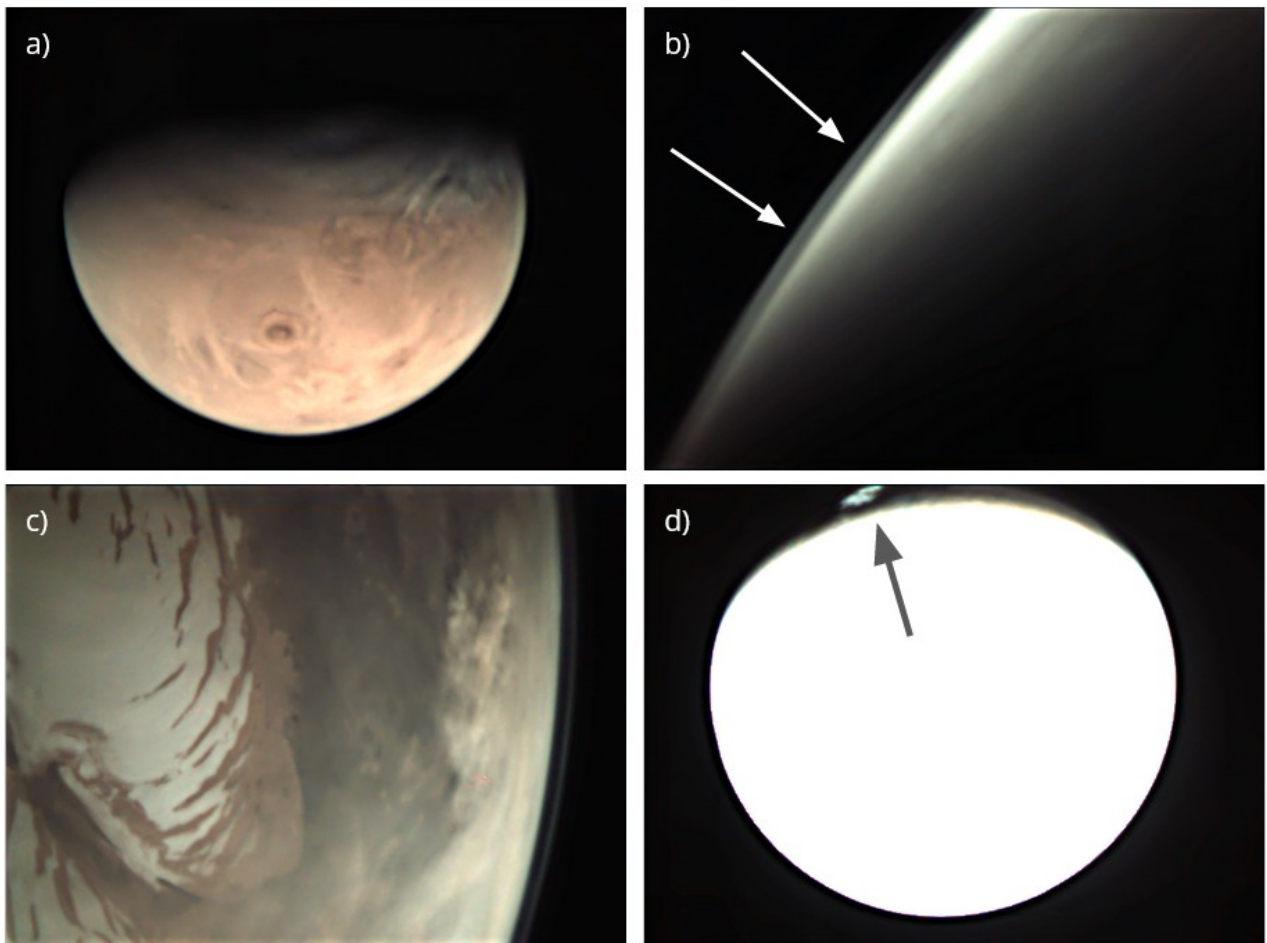

Fig. 2. Examples of different VMC observations. (a) Full disk image of Mars from apocenter. (b) Aerosols in the limb of Mars imaged in high resolution during a near-pericenter observation. (c) North Polar Cap and a dust storm observed in high resolution near the pericenter. (d) A cloud in twilight seen in an intentionally overexposed full-disk image from apocenter.

VMC has a wide FOV. Together with the elliptical orbit of MEX, this enables full-disk observations from apocenters (Fig. 2a). This capability has motivated new studies of a seasonal cyclone regularly taking place in the northern planitiae (Sánchez-Lavega et al., 2018b) and of textured dust storms regularly appearing around the north polar cap during the first half of the year (Sánchez-Lavega et al., 2022a; 2022b; dust storm example in Fig. 2c). The apocenter position during the recent Global Dust Storm in 2018 allowed a regular monitoring of the south polar region, which was presented in Hernández-Bernal et al. (2019). The instrument's wide FOV can also be used to image large sections of the limb in search for high altitude aerosols (Fig. 2b), as shown by Sánchez-Lavega et al. (2018). In addition, taking advantage of this wide FOV, some test observations were performed trying to find meteors in the atmosphere; such use of orbital cameras with wide FOVs was proposed by Christou et al. (2012). However, the low sensitivity of the sensor and the long time between images (limiting the effective integration time) limit de detection capabilities of VMC for this purpose, with unsuccessful results in the observational campaigns executed (Hernández-Bernal et al., 2021c).

In addition, MEX allows the imaging of a wide range of local times, whereas some other missions are in sun-synchronous orbits where only a narrow range of local times are reachable. This capability, together with VMC's full-disk imaging, has been key to the study of a rare elongated cloud on Arsia Mons that has been observed in the early morning hours (Hernández-Bernal et al., 2021a). It has also allowed novel studies of aerosols (Hernández-Bernal et al., 2021b; Fig. 2d displays an example of such aerosols).

The third standout capability of VMC is the possibility of obtaining sequences of images capturing the evolution of dynamic events in the atmosphere over several minutes. This is possible because of the small image size that enables the acquisition of several images without a significant impact to the memory and downlink budgets, and also because of the wide FOV that enhances the chances of capturing dynamic events for several minutes. From these videos it is possible to obtain measurements of motions in clouds and dust features, enabling the direct measurement of winds, which is scarce type of measurements on studies of Mars. This VMC capability has been exploited by Sánchez-Lavega et al. (2018b) and Hernández-Bernal et al. (2019; 2021a).

A general review of dynamical phenomena in the atmosphere of Mars based on visible images can be found in Sánchez-Lavega et al. (2024), with VMC taking a leading role in many of the investigations referenced there.

### 3. VMC Science Operations

Before May 2018, VMC was operated directly by the engineering teams in the Mission Operations Centre at ESOC (Germany), independently of the science instruments on MEX. As shown in Fig. 3, the number of images in the first Martian Years (MY28-MY33) was very low, with only a few exceptions for dedicated VMC observation campaigns[2].

Starting in May 2018, corresponding to Solar Longitude (Ls) ~180º in Martian Year 34 (MY34, see Fig. 3), VMC was included in the payload science operations timeline and harmonized with other instruments by the Science Ground Segment (SGS) at ESAC (Spain), based on the scientific objectives and priorities defined by the science team. The new science observation scheduling (described in this section) resulted in a significant increase in the number of VMC images producing tens of thousands of images from mid-MY34 throughout MYs 35-37 (Fig. 3). VMC operations are still constrained by the hardware limitation that other instruments cannot observe when VMC is switched on, the 19º tilted boresight compared to other instruments, and the principle of not interfering with observations by other scientific instruments except in exceptional circumstances.

This section describes the operations of VMC through the three stages of MEX Science Operations as defined by the SGS (described for example in Cardesín-Moinelo et al., 2024). Once images have been acquired, the uncompressed data is downloaded to the ground within hours or days.

---

2  e.g. VMC movies in May 2010, Solar Longitude ~90º in MY30
   https://www.esa.int/ESA_Multimedia/Videos/2010/06/Full_orbit_How_an_astronaut_will_view_Mars_from_orbit

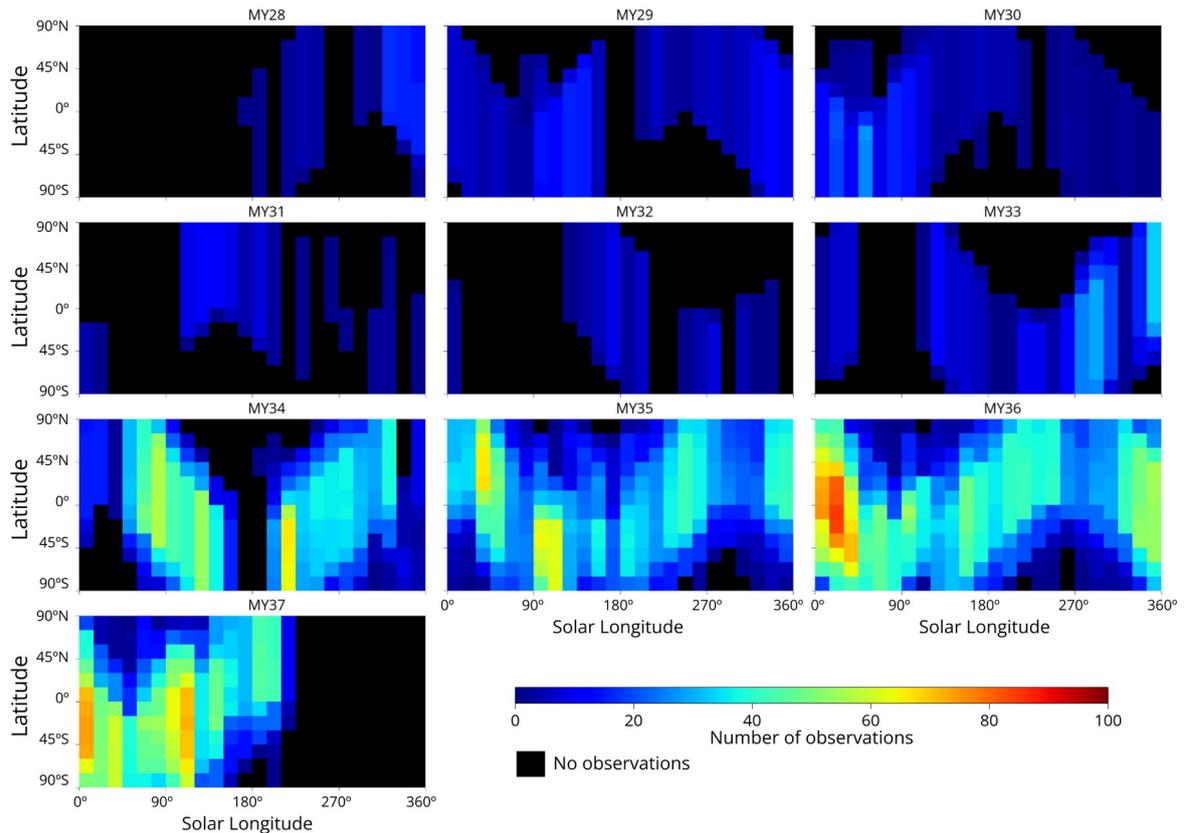

Fig. 3. Coverage achieved by VMC over time, up to March 2024. Each panel is a different Martian Year (MY). Horizontal axes correspond to the Solar Longitude (time of the Martian Year). Vertical axes correspond to latitude.

### 3.1. Long Term Planning (LTP): Science Opportunity Analysis

The MEX nearly polar orbit precesses slowly changing the latitude of the pericenter and apocenter and the right ascension of the orbital nodes. Thus, the phase angle at which observations of the planet are acquired can change from very high to very low angles on different phases of the mission (dayside/nightside) as shown in Cardesín-Moinelo et al. (2024). During this long-term planning cycle, the VMC and MEX SGS teams perform a science opportunity analysis using the mission reference trajectory for the coming months/years, computing the visibility and illumination conditions of certain regions and targets of interest (e.g. the polar caps, Tharsis, spacecraft landing sites, etc.) over long periods of time. This serves to identify in advance the main observation campaigns and define the global scientific priorities for the upcoming seasons.

In particular for VMC, special attention is given to the evolution of the apocenter latitude and illumination conditions. High altitude apocenter observations are particularly useful as Mars is far enough away (~10,000 km) to fit inside the VMC FOV, enabling the science team to monitor the global atmosphere on a regular basis. Given the characteristics of the MEX orbit, the apocenter can be in the dayside or nightside of the planet with varying latitudes, although the long-term analysis shows a bias around low latitudes towards the morning side of Mars, so the evening can only be observed from apocenter when it is over polar latitudes. A nice example of how this long-term evolution biases the local time coverage is shown in Fig. S2 in Hernández-Bernal et al. (2021b).

### 3.2. Medium Term Planning (MTP): Observation Scheduling

During the MTP, the timeline of operations for each orbit is defined, the observation time is distributed between instruments and the pointings of the spacecraft are defined. Due to hardware limitations when VMC is switched on, other instruments cannot send data to the mass storage, and thus their data is lost. While most MEX instruments only observe in particular windows, this is a problem for ASPERA, which is a plasma monitoring experiment that ideally should always be observing (Barabash et al., 2004).

As a result, most VMC observations take place around MEX "maintenance blocks", performed roughly once a day at apocenter, in which MEX turns on its engines to unload the reaction wheels, maintain the orbit trajectory and perform other essential operations. ASPERA cannot observe during maintenance apocenters because particles being expelled by MEX thrusters would affect the results. With the new science planning scheduling implemented in 2018, VMC observations are routinely planned whenever there is a suitable window, a few minutes before and after these maintenance apocenters, with the maximum number of images possible (typically 8 before and 38 after).

In addition to the apocenter observations, VMC is allowed to observe at other locations of the orbit typically once a week or more frequently for special campaigns, in agreement with the ASPERA team. This allows for the routine planning of low altitude observations near pericenter, that are dedicated to covering various science objectives. These include: the tracking of clouds at the limb (Sánchez-Lavega et al., 2018a), tracking specific target regions at different local times as done during the observation campaign for the Arsia Mons Elongated Cloud (Hernández-Bernal et al., 2021a), following the aerosol evolution around the polar caps (Sánchez-Lavega et al., 2018b; Hernández-Bernal et al., 2019, Río-Gaztelurrutia et al., 2021) and monitoring regional meteorology over certain landing sites for collaboration with surface landers and rovers (Sánchez-Lavega et al., 2022b).

Moreover, in some particular seasons ASPERA cannot observe continuously, due to thermal, power or data volume restrictions, and this provides VMC with opportunities to perform very long and regular observation campaigns, extending the global coverage over certain seasons, as is visible in Figure 3 at the beginning of MY36.

### 3.3. Short Term Planning (STP): Commanding

For each MTP, there are STP phases that run closer to the actual operations. During the STP, each instrument defines the detailed commanding and observational parameters, which in the case of VMC, involves setting the exposure times.

### 3.3.1. VMC Image Loop Acquisition

Until 2012, VMC images were commanded individually with variable configuration settings, but this required too many commands and this approach was no longer possible given the way MEX currently operates. In 2012, the VMC commanding was updated, and now each VMC observation consists of a series of consecutive images with the same exposures that are obtained in a loop of 6 images that are repeated several times. When the instrument is switched on, it first takes a "default" image with an exposure of 14 ms. Then, a sequence of 6 images is obtained and the following images follow the same sequence of 6 images acquired with the same exposure times. The loop can be repeated several times. Fig. 4 illustrates a typical observation sequence with a timeline of the full observation sequence. Note that between 2012 and 2018, the length of the loop changed occasionally, but no more updates are expected in the future.

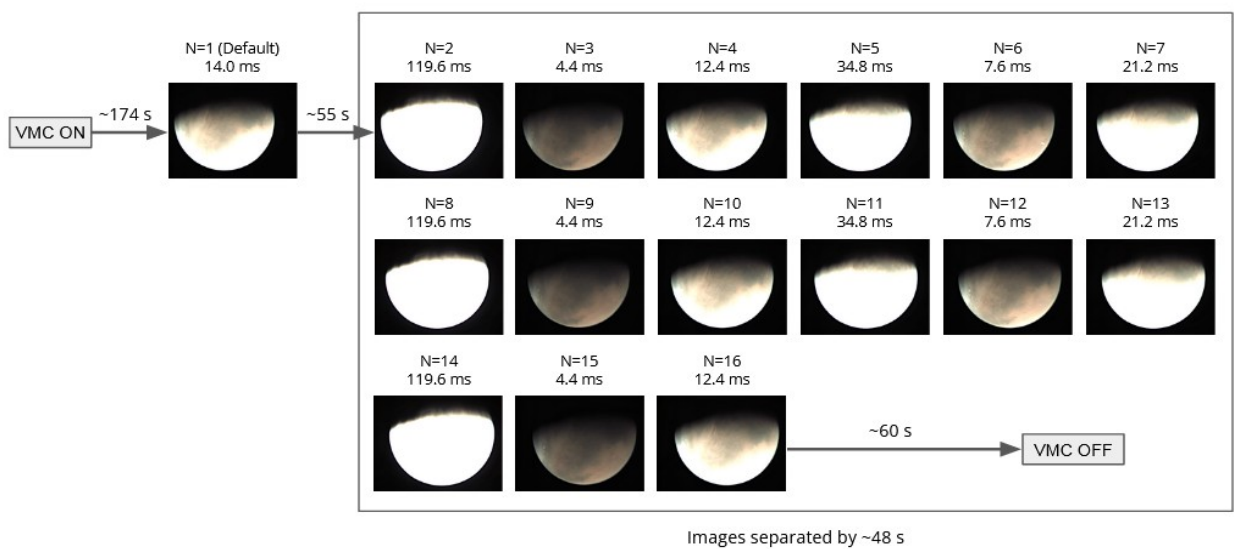

Fig. 4. Schematic image loop acquisition of VMC.

The exposure of the initial "default" image of 14 ms is hardwired and cannot be changed. The 6-image loop parameters are configurable and are commanded based on a specially designed strategy detailed below in section 3.3.2.

As mentioned in section 2.1, VMC requires 38 seconds for each image to be transferred to the memory, and an additional margin that can be partially configured for each image of the loop is added. The time between images is typically around 48 seconds. An exception to this is the interval between the first (default) and the second image, that is typically around 55 seconds, and the last image before switch OFF, which is typically a couple of seconds faster. Additionally, 174 seconds are required to switch on VMC, and up to 60 seconds are required to switch it off. This is also illustrated in Fig. 4.

### 3.3.2. Setting exposure times
The exposure time can range from 0.4 ms to around 90 s. The lower exposure mode, typically used for science observations, ranges from 0.4 to 191.6 ms, in steps of 0.8 ms. The higher exposure mode is used mostly for calibrations using stars, and it suffers from saturation by thermal noise, which typically overflows the whole sensor even for dark sky observations when the exposure is longer than ~40 s (depending on the temperature of the sensor). The borders of the image field can become saturated with exposures of 20 s.

As VMC images often cover the whole disk of the planet, a wide variety of incidence, emission, and scattering angles are present, and this is a challenge when optimizing exposure times for regular observations. Typical exposures for the dayside of Mars are around 2-20 ms, but higher exposures (20 - 120 ms) are usually set to see thin aerosols over the limb (Sánchez-Lavega et al., 2018a) or in the night side of the terminator (Hernández-Bernal et al., 2021b). Even higher exposures, of up to 20 - 30 seconds have occasionally been used to image stars for calibrations (see section 5.1) or for particular experiments (e.g. Hernández-Bernal et al., 2021c).

The dynamic range of VMC is limited by the 8 bits used to code the images (see Table 1). This implies that, within a single image, there is a trade-off between overexposing bright areas (high exposures) and capturing too much noise on dark areas (low exposures). In order to overcome this constraint, we use sequences of images with different exposure times to observe both bright targets (e.g. features in the dayside of Mars) and dark targets (e.g. dim aerosols high on the limb or far into the night). The strategy for setting the specific sequences of exposure times within VMC loops is oriented to optimize the production of high-level products in the form of HDR images (section 4.3), while keeping a high temporal resolution, which also enables high level products in the form of videos showing atmospheric dynamics.

## 4. VMC Science Data

Both VMC raw and calibrated science data are regularly produced and uploaded to the ESA Planetary Science Archive (PSA). The technical information describing the archived data is given in the VMC EAICD (Experiment to Archive Interface Control Document) and in Ravanis et al. (2019; 2020). Here we concentrate on providing details of the science calibration procedure (section 4.1), describing artifacts remaining after calibration (section 4.2), and detailing our procedure to build high level HDR images (section 4.3).

### 4.1. Photometric Calibration of VMC images

We calibrate raw images following the standard procedure of subtracting a dark current and dividing by a flat-field. In our case, the flat-field correction is much more relevant than the dark current correction in the quality of the final images after calibration. The resulting images can be used for relative photometry. However, obtaining absolute photometry data is challenging given the difficulties in characterizing the spectral response of the Bayer mask covering the VMC sensor, and thus, is not covered here.

Saturated pixels (i.e. with a value of 255) are converted to NaN (Not a Number) at the start of the calibration, so that they are not handled as if their value were physically realistic.

Given that our photometry is relative and not absolute, calibrated images are stored in arbitrary units resulting from this procedure, and the values of pixels from different images need to be divided by their exposure time to be comparable.

The following subsections describe the bias/dark current and the flat-field of VMC and the procedures used to generate such corrections. So far, the long-term evolution of these corrections has not been investigated, and this might be a subject of future updates in the calibration pipeline, which would be documented in the VMC EAICD.

### 4.1.1. Bias and Dark

Here we describe the most general case of low exposure images (low exposure mode reaches up to 191.6 ms, see section 3.3.2). Higher exposures are only used for calibrations, and that case is briefly described in section 5.1. We base our analysis on images acquired by pointing VMC to the dark sky, specifically to a region around the constellation of Eridanus with no stars brighter than magnitude 3 over an area with an angular size comparable to the VMC FOV. Given the low sensitivity of VMC, the brightness of stars in this region selected is negligible in the low exposures used. The angle from the VMC boresight to Mars and to the Sun was ~100º and ~140º respectively, which prevents light from those sources from reaching the sensor.

In low exposure mode, the background noise exhibits typical values of DN≤3, with higher values only in the uppermost line. No significant dependency on temperature has been observed. The exact distribution of the background noise on the sensor is a function of the exposure time, as displayed in panels a-c in fig. 5. In each case, the image can be divided in two horizontal sectors separated at $y_0=2.5 \cdot \text{exposure\_time}$. Since exposure time ranges from 0.4 to 191.6 ms in steps of 0.8 ms (section 3.3.2), this formula produces integer odd values ranging from 1 to 479, covering the whole vertical extension of the sensor. Each one of the two horizontal sectors displays three levels of noise (with DN=0, 1, and 3) increasing from left to right at positions that are different in the upper and the lower sector. The distribution of noise is slightly different in the default image (see section 3.3.1; panel d), but the levels of noise are similar.

Given the low level of background noise (DN≤3), especially when compared to the typical values of the glare observed in images showing Mars and described in section 4.1.2, we choose to follow a simple approach for the bias and dark current correction that does not consider all these details. Instead, we simply apply a correction with a noise distribution corresponding to a low exposure image with an exposure of 6.8 ms.

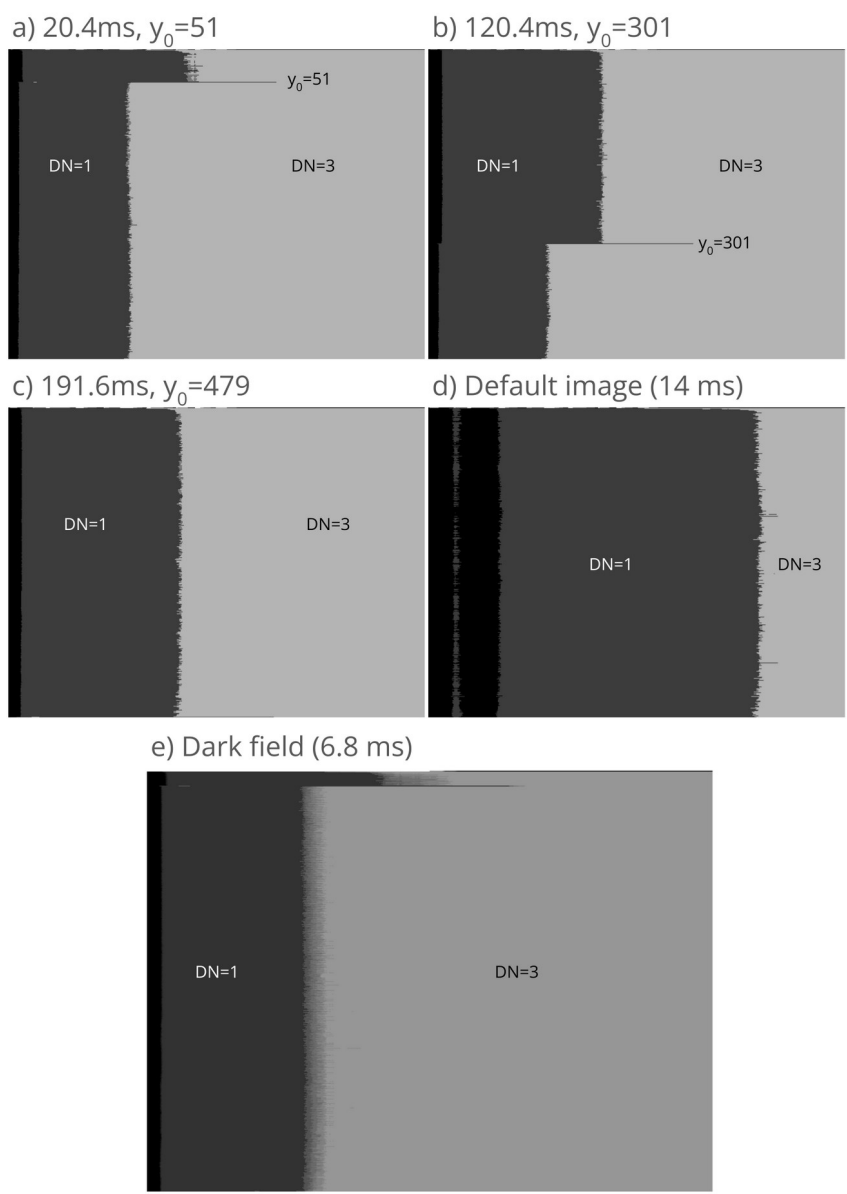

Fig. 5. Distribution of background signal as a function of exposure time (panels a-c) and for the default image (panel d). Dark field employed for calibration (panel e), with a distribution of the dark current typical of an exposure of 6.8 ms.

### 4.1.2. Flat-field

The flat-field was created using dark-corrected images of flat portions of Mars that were well and uniformly illuminated and as free as possible from large structures, so that each individual image had an overall distribution of its illumination as flat as possible. Suitable images were obtained from MEX pericenter with VMC looking to the northern plains of Mars. The flat-field was built combining 120 images selected from several sequences of observations. The list of these images is given in the supplementary material. The images were combined using sigma-clipping, a technique that rejected pixels whose brightness value were two standard deviations away from the mean value of the pixels across all the images for darker images, and 4 standard deviations away for higher pixel value images. This different treatment of dark and bright images was incorporated so that images with poorer illumination had less influence on the final master flat. Once this sigma-clipping was performed, the remaining pixels were averaged across all the images and then normalized to create the master flat-field file with a mean brightness value of 1.0. The mean standard deviation of this averaging was around 10%.

The resulting normalized flat-field is shown in Fig. 6a. The maximum value is 2.1, and the lowest value is 0.02; but as we show in the histogram in Fig. 6b, most pixels take values between 0.4 and 1.4. We note that it is possible to separate two clearly different components in the flat: the vignetting component (Fig. 6c), and the noisy component (Fig. 6d). We will comment on the features of the flat field in section 4.2.

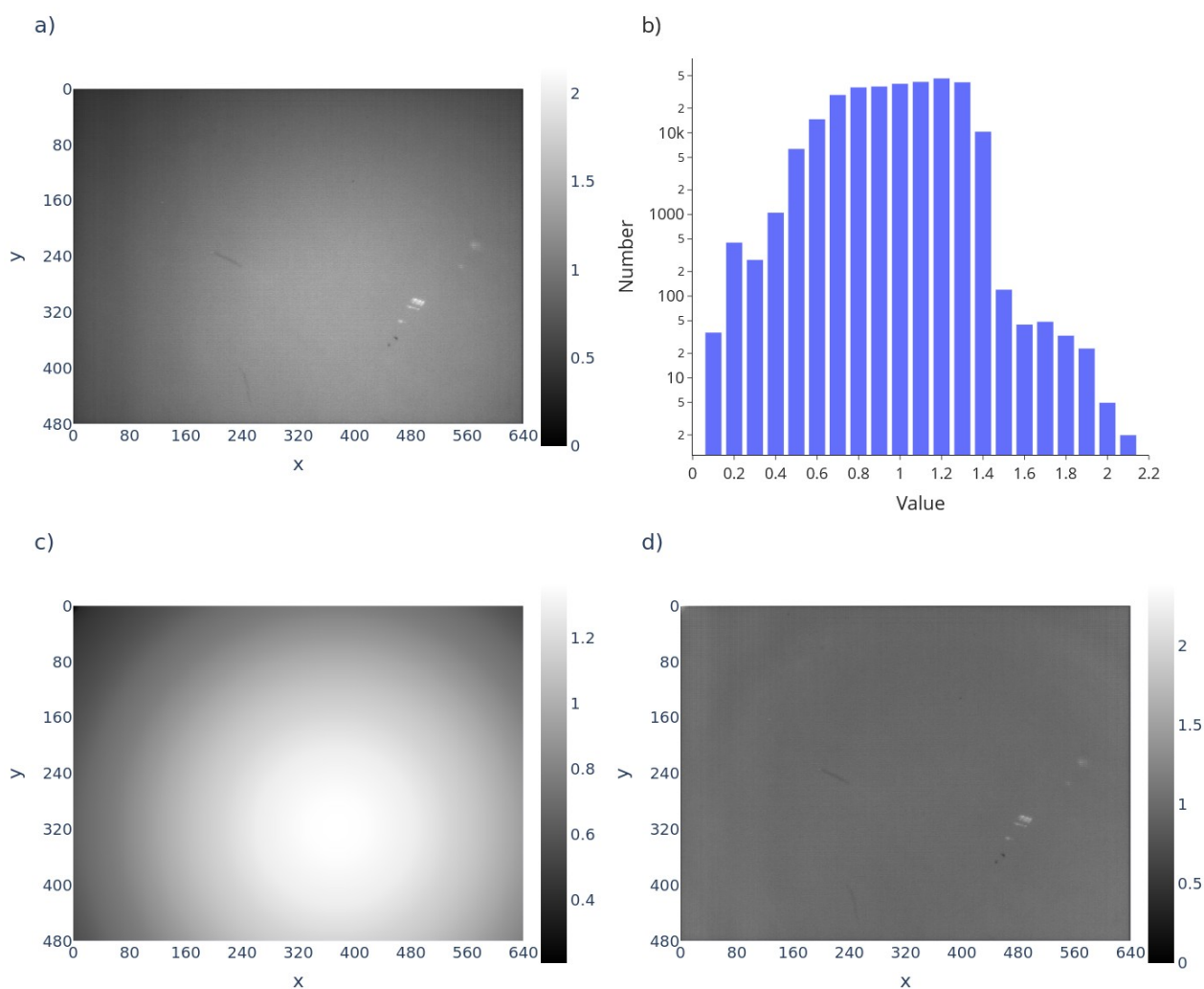

Fig. 6. Flat field and details about its structure. Values of the histogram and color scales correspond to the normalized flat (a) VMC flat field. (b) Histogram in log-scale showing the frequency of different values of the flat. (c) Vignetting component of the flat. (d) Noisy component of the flat.

## 4.2. Main known artifacts in VMC images

There are some artifacts and issues on the VMC sensor. Many of them are reasonably well corrected with the application of the flat field. Others are more complex and not well understood. Some artifacts have not been investigated in depth because they do not impose limitations on current VMC science investigations. We describe below the artifacts that in our experience should be taken into account during the analysis of VMC data:

- The "Beagle Regards" (Fig. 7a). This is a prominent artifact in the middle right area of the sensor (centered around x~500, y~ 300, with a size of the order of 100 pixels. See detailed distribution in fig 7a) that was initially thought to be a defect created during the release of the Beagle-2 lander. However, it is already present in images acquired by June 2003, months before the separation of the lander, and its actual nature is unknown. In most images, the flat field correction solves the effects caused by this artifact reasonably well. However, when the pixel in original raw images have high values, some reddish pixels remain even after the flat is applied. This artifact is also present in dark images with long exposures (a few seconds and more). As a consequence of this artifact, faint structures in this part of the sensor have to be looked at carefully; Hernández-Bernal et al (2021b) detail in their supporting material the need to implement an exception in two pixels of the field due to this artifact (points given by x,y=460,358; and x,y=450,367, as given in that reference).

- Middle left dark line (see its location and form in fig. 7b). This is a dark feature in the flat field that can occasionally be confused with something real. It is successfully removed by the flat field correction during calibration.
- Top lines (Fig. 7c). Some defects appear in horizontal lines in the top of the sensor. Their specific position is usually around the $y_0$ line (which is a function of exposure time, as described in section 4.1.1). They have not been investigated in detail and the current calibration process is not robust enough to remove them.
- Glare (Fig. 7c). Images with the full-disk of Mars are always affected by a glare that is not well understood. Its characteristics are presented in detail below, in section 4.2.1.

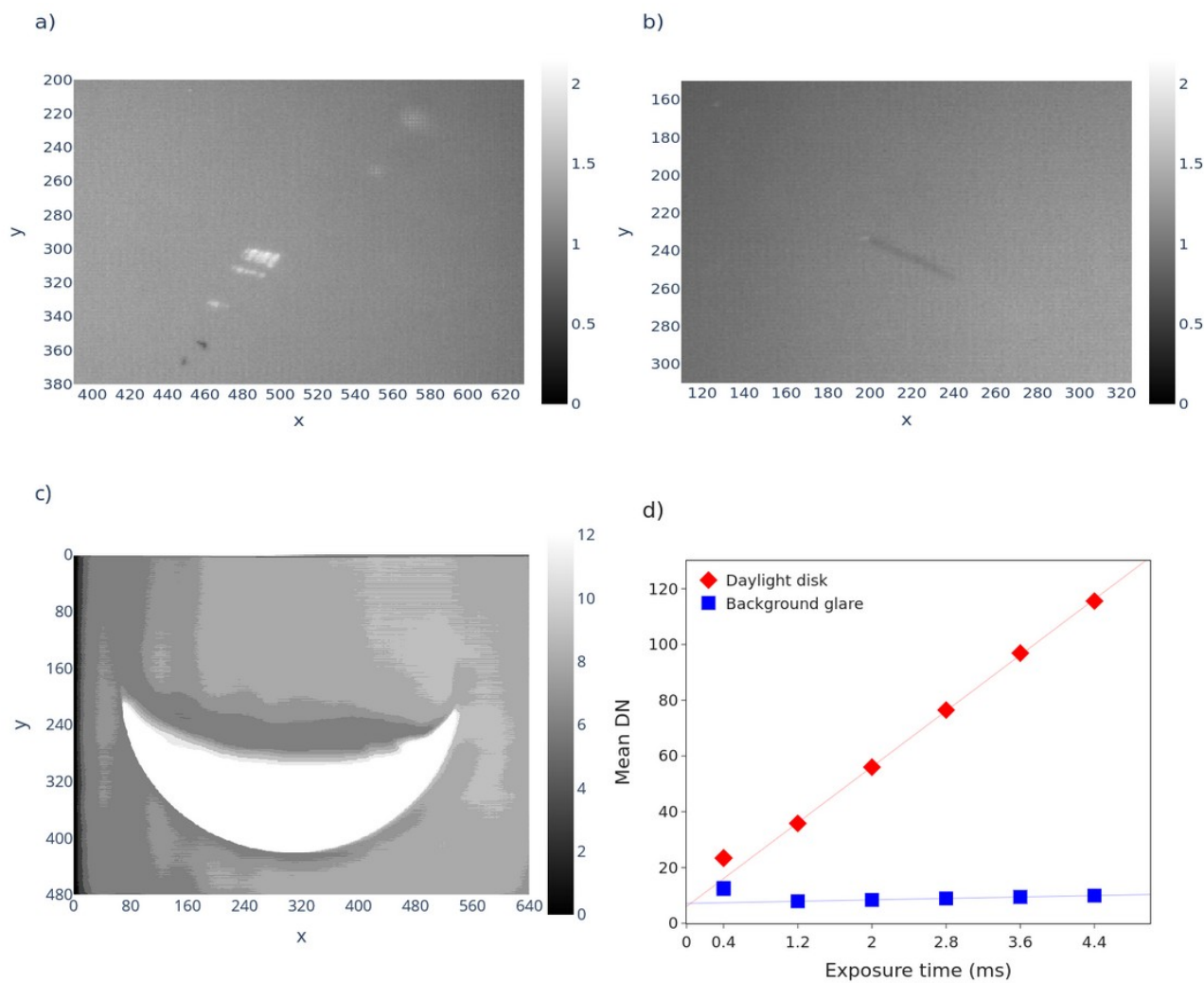

Fig. 7. Main known artifacts and issues in VMC images. (a) "Beagle Regards" as seen in the flat field (same as Fig. 6a but zoomed on the artifact). (b) Middle left dark line as seen in the flat field (same as Fig. 6a but zoomed on the artifact). (c) An image from observation MEXVMC_200292, which was a test for the glare. The color scale limits have been adjusted to the range DN=0-12 to highlight the glare brightness around the disk of Mars. The artifact in the top lines of the sensor is also apparent. (d) Analysis of the brightness in the same observation. Red points represent the mean brightness of the illuminated side of the disk of Mars as a function of exposure time. Blue points represent the mean brightness of non-illuminated pixels, which corresponds to the background glare and is almost constant with exposure time.

### 4.2.1. The VMC glare

The glare consists of a background noise whose intensity is not proportional to exposure time and that only appears when a minimum amount of light reaches the sensor. The glare does not appear in dark images (section 4.1.1) and does not appear in images with faint sources of light (e.g. Phobos). However, this glare appears even when a faintly illuminated section of Mars appears in a corner of a VMC image and affects the whole sensor whenever the glare appears. Therefore, all images showing the dayside of Mars are affected by this artifact. An example of this glare is displayed in fig. 7c.

The brightness distribution of this glare varies in images with different distributions of the input light (i.e. different shapes and positions of the illuminated part of the disk of Mars). While we have developed significant efforts to try to understand this glare and develop a correction for it, the conclusion for now is that it cannot be easily modeled or removed from the images. We describe here the relevant results obtained from our analysis.

We ran a specific observation (VMC observation MEXVMC_200292) to check the linearity of the glare with respect to exposure time. We used a sequence of low exposures from 0.4 ms to 4.4 ms with intervals of 0.8 ms and measured the brightness averaging DNs of pixels inside and outside the illuminated part of the disk of Mars (red and blue dots in Fig. 7d respectively). Then we calculated the y-intercept of the different exposures (i.e. the expected brightness to be recorded with a null exposure). We found that the mean value of pixels outside the daylight disk of Mars is almost constant with exposure time (blue points in Fig. 7d), which suggests that the glare is mostly of electronic origin and not an artifact introduced by the optics. For the lowest exposure of 0.4 ms, the glare was brighter everywhere (Fig. 7d) and had a visible effect over the disk of Mars, which was poorly illuminated due to the low exposure. However, for longer exposures, the average glare exhibited a lower and almost constant brightness out of the disk of Mars, and the brightness of the daylight side of the disk of Mars grew linearly with exposure, with a y-intercept of ~6 DNs originated by the glare.

We conclude that for well illuminated areas the effect of the glare is low (6 DNs on average), which might impose limitations for accurate photometry. Analysis of dark features in the limb or the terminator should be performed carefully for poorly illuminated features, which might be affected by the glare. Previously published works analyzing dark features at the limb and the night side of the terminator (Sánchez-Lavega et al., 2018; Hernández-Bernal et al., 2021b) discarded from their analysis unclear features that could be related to the glare.

## 4.3. Production of High Dynamic Range (HDR) images

VMC calibrated images can be combined and stacked together to obtain high level HDR images. The combination and stacking is done after projection of all source images to a common geometry, which avoids imperfections due to the rotation of the planet and the movement of the spacecraft. This procedure has been successfully applied to VMC works already published (Hernández-Bernal et al., 2019; 2021a), and is illustrated in Figs. 8 and 9.

The workflow to build HDR products from a group of VMC images involves the following steps:
1. Select source images and project them to a common geometry (whether a projection on a sphere observed from a common point of view or a projection into a planisphere).
2. Normalize the brightness of the source projected images by dividing their values by their exposure times.
3. Map the incidence angle in the destination projection, assuming a timestamp that is typically the average of the timestamps of all the source images.
4. Define regions of optimal exposure times defined by a range of incidence angles and luminosities in the original images (see Fig. 9).
5. Build the final image from different combinations of the original images in each different region (see Fig. 9). The pixels in a region of the final image may or may not be associated with a combination of different exposure times, depending on the convenience of stacking different exposures in a single region and considering the specific illumination and available exposures for that particular region of the image.

These steps result in a product that conserves the relative photometry and contains more information than what can be stored with a color depth of 8 bits, all calculations and storage are done in single-precision floating-point. Tone mapping needs to be applied to export this HDR product to an 8 bits image in which most areas look properly illuminated, then the relative photometry is lost. In the case of planetary images, it makes sense to use typical luminosity corrections (e.g. Lambert cosine law) for tone mapping. Right panel in fig. 8 is an example of HDR image after tone mapping.

Geometrical calculations and projections are made with the software "Elkano", which was first described in (Hernández-Bernal et al., 2019) and will be further documented elsewhere.

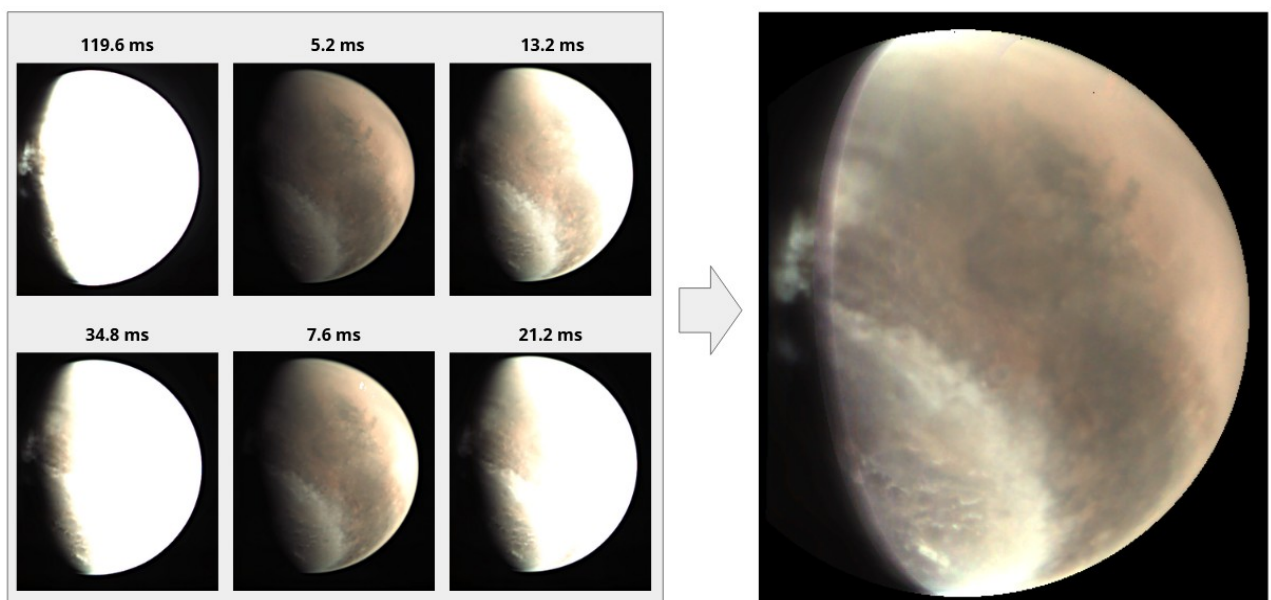

Fig. 8. Example of a HDR image (right) made from a loop of 6 images with different exposure times (left). A luminosity correction (tone mapping) was applied to the HDR image to display it as standard 8 bits image. These images were obtained on 2022-01-10.

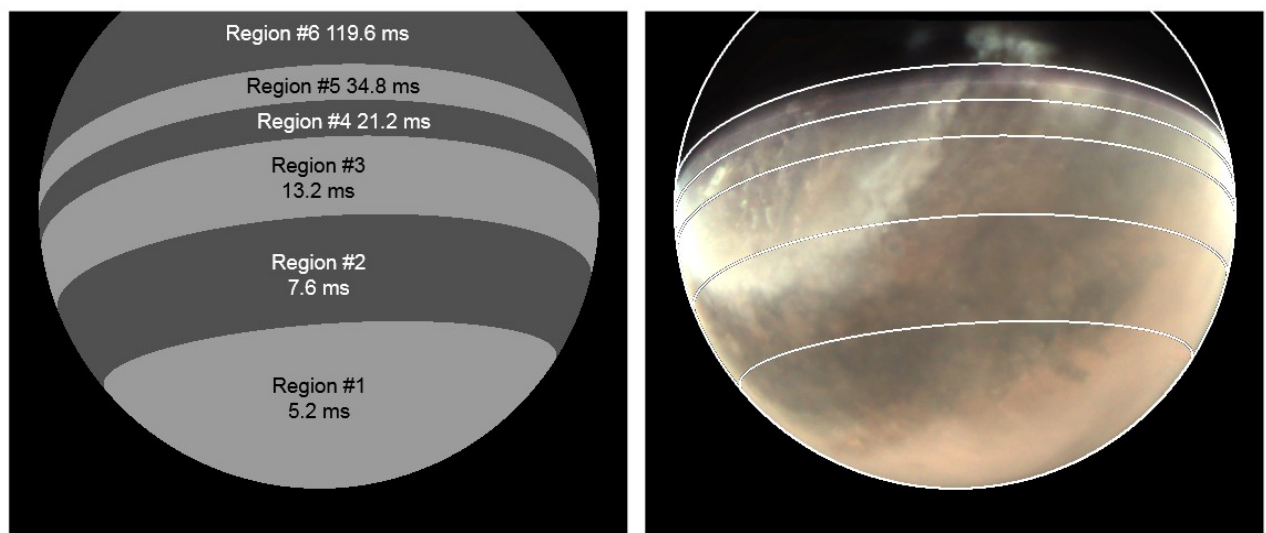

Fig. 9. Regions in a destination projection during the process to make an HDR product. The left image is the scheme of the destination projection, where different regions with different optimal exposure times (from 5.2 to 119.6 ms) have been determined, each one associated with a single optimal exposure time. The right image is the output image with the different regions indicated. A version of this image, rotated 90º counterclockwise and without marks can be seen in the right panel of Fig. 8.

## 5. VMC Geometrical Calibration

The determination of the exact geometry of the images requires an accurate knowledge of the camera, and its position and orientation relative to the planet. The Flight User Manual indicates the design parameters for the orientation of VMC in the reference frame of MEX, and the pixel size and shape. However, the accurate parameters once VMC was mounted on MEX were never measured on the ground.

In addition to this, we found that the timestamp of images suffers a random shift of a few seconds; the timestamp is registered after the actual image is captured, and since VMC was not designed for scientific accuracy, this results in new uncertainties.

As a result, we had 5 free parameters that were only constrained by their values in the Flight User Manual and needed to be determined:
- The 3 Euler angles (also expressed as a rotation matrix) that determine the accurate orientation of VMC relative to the MEX reference frame.
- The pixel instantaneous Field of View (iFOV), which defines the pixel angular size and shape. The pixels might not be a perfect square and there might be optical aberrations in the optics. From our analysis of VMC images, we can rule out that optical aberrations or pixel irregularities play a significant role and we can consider a fixed and unique value for the pixel size.
- The shift in time from the actual image acquisition timestamp to the labeled timestamp. Our analysis shows that the assignment of timestamps suffers random variations that prevent a perfect solution for this problem.

We explain our determination of the VMC rotation matrix and pixel size in subsection 5.1. The analysis of the shift in time is explained in subsection 5.2.

## 5.1. Determination of VMC orientation and FOV

In order to determine the orientation of VMC relative to the Mars Express reference frame, we acquired images of star fields that could resolve this orientation with a subpixel precision. Several star observation campaigns were executed in 2019 and 2020 pointing VMC to the sky and using exposure times of around 20-30 s. Such long exposures require a special treatment as the low exposure bias presented in section 3 is not valid anymore. For these observations we used dark current corrections derived from images obtained with similar exposure times. From the analysis of the images we determined that the VMC signal to noise ratio in this kind of observation allowed us to see stars of magnitude up to 2-3. Many of the star observations covered the constellation of Orion and additional tests were done observing Jupiter and Saturn treated as stars. During these observations the spacecraft maintained a fixed attitude, with an accuracy much higher than the VMC iFOV. Therefore, the time related uncertainty is not present and only the 3 Euler angles and the pixel size remain as free parameters.

The Flight User Manual states that VMC is tilted 19º around the spacecraft X axis (panel of the High Gain Antenna), that is the initially known information about the VMC reference frame. It also indicates that the pixel size of the sensor is $s=14$ μm (see table 1), and the focal length of the optics is $f=12.3$ mm, which corresponds to a pixel angular size of: $iFOV=s/f=1.13821$ mrad.

We processed these images to extract clearly visible stars and removing noise. The position of each star in the detector was determined by a gaussian fit to a subpixel accuracy. Due to noise in the VMC sensor, some stars have a point-spread-function in which the gaussian fit can be inaccurate. To reduce this problem, the gaussian fit was compared to positions of the stars in the detector determined manually. When the difference between the manually determined position and the gaussian fit is higher than 2 pixels the star was discarded from the analysis. The result was a set of 54 stars covering most of the area of the VMC sensor. The observed position of stars and an example image showing the constellation of Orion is shown in Fig. 10.

We then tested a large set of combinations of the 3 Euler angles and pixel size using a brute-force algorithm that examined the four dimensional space of parameters for the different observations of star fields fitting all the observations simultaneously. The best result, which fits the position of 46 stars with an accuracy of less than 1 pixel, is given in Table 2.
The calibrated values are available in the updated VMC Instrument Kernel (file name: MEX_VMC_V04.TI) as part of the Mars Express SPICE kernel dataset, that can be opened and used for geometric calculations using the SPICE software (Acton et al., 1996; 2018).

Table 2. VMC Geometry parameters

|  | *Value from Flight User Manual* | *Calibrated value* |
|---|---|---|
| $\alpha_x$ | 19º | 19.2434º |
| $\alpha_y$ | 0º | 0.2743º |
| $\alpha_z$ | 0º | 0.4251º |
| iFOV | 1.13821 mrad/px | 1.12977 mrad/px |
| FOV | 40.03º x 30.56º | 39.75º x 30.34º |

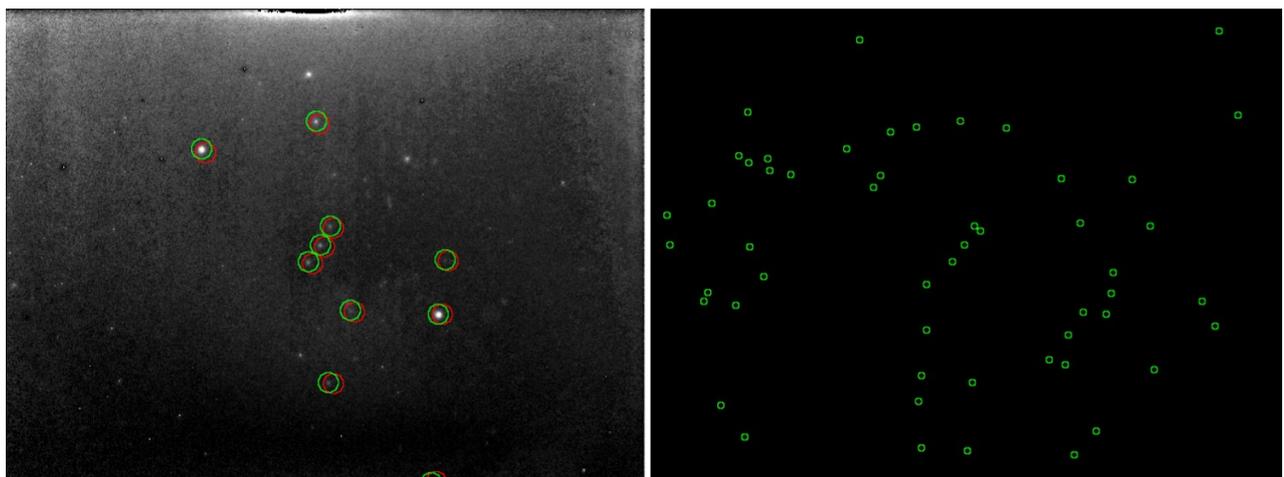

Fig. 10. Geometric calibration of VMC. (left) Example of a calibrated image showing the constellation of Orion, the expected position of stars is indicated with red circles (geometry based on the Flight User Manual), and green circles (calibrated geometry). (right) Positions on the sensor of 54 stars used in the analysis once the different fields of stars have been combined.

## 5.2. Time: Using Phobos as a "clock"

The remaining free parameter is the precise time of the observation around the timestamp available for each image. In order to determine the accurate time of individual images and compare it to the registered timestamp, we used observations of Phobos whose position acts as a "clock". Phobos orbits Mars with a nearly circular orbit with a semi-major axis of 9,376 km and an orbital period of 7h 39m 12s. Mars Express apocenters are ~10,000 km and Phobos is sometimes present in VMC observations from apocenter. In such cases, Phobos moves quickly across the VMC FOV. The shadows of Phobos and Deimos on Mars offer alternative "clocks". However, the shadows of Phobos and Deimos over Mars are blurry because these satellites do not produce full eclipses over the surface, and thus, their position is difficult to determine accurately in VMC images. In addition, these shadows move more slowly over the sensor which limited their usage in this investigation, and so they were ultimately not considered in the analysis.

We used 37 individual images showing Phobos, with some of them having Phobos transiting the disk of Mars. The observed position of Phobos was manually indicated in each image, and then a specially made program tested different timestamps a few seconds before the labeled timestamp, determining the actual timestamp for each individual image showing Phobos. This is illustrated in Fig. 11.

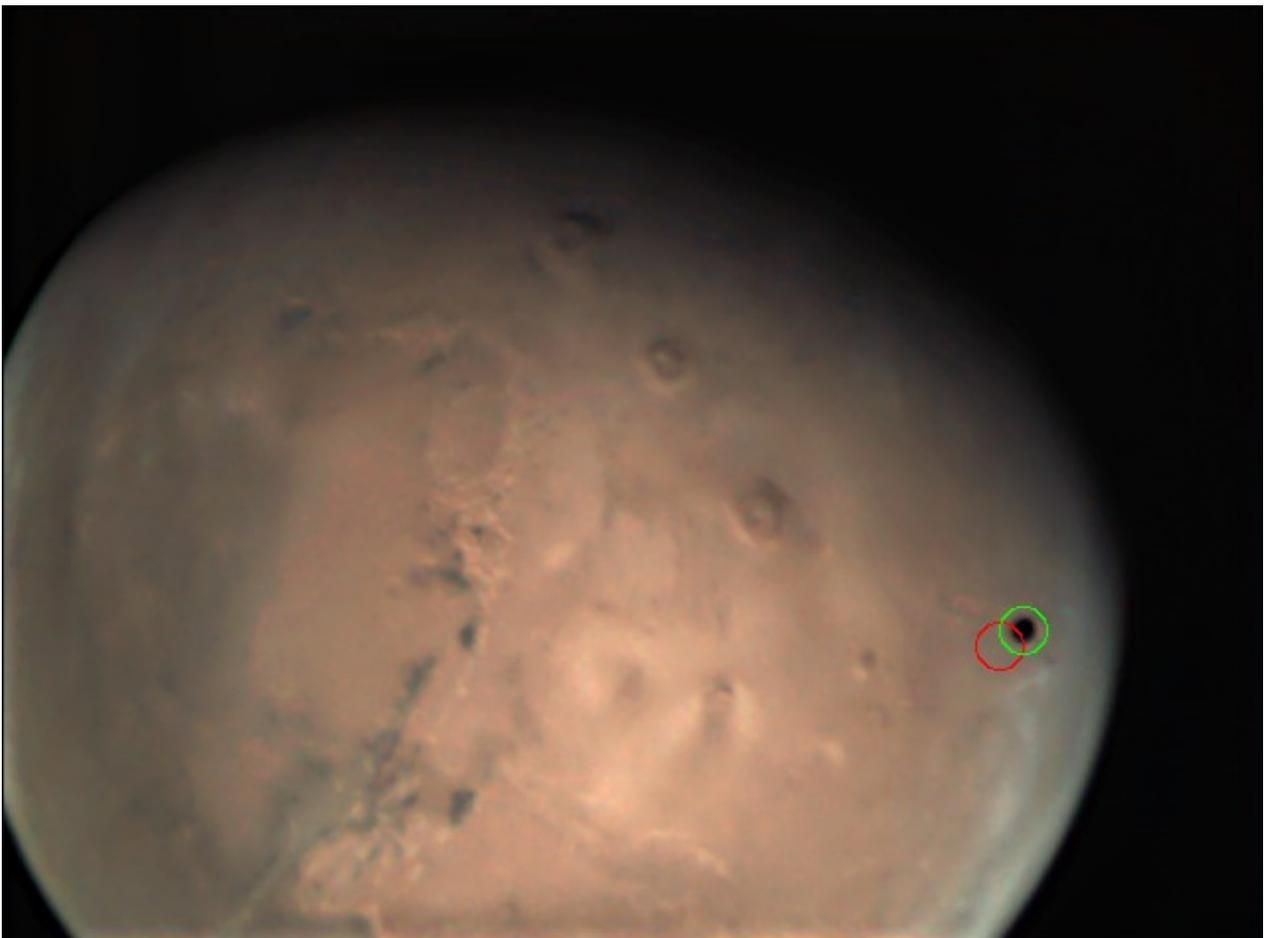

Fig. 11. Phobos transiting the disk of Mars as seen by VMC on 8 February 2021. The registered timestamp was 18:58:11, but the expected position of Phobos at this time (red circle) is different from the observed position, which is reproduced more accurately by subtracting 10 seconds (timestamp 18:58:01; green circle).

The fitted timestamps were between 6 and 13 seconds before the registered timestamp, with a mean shift of 9-10 seconds. However, the specific shift for each image seems to change randomly and no clear pattern can be extracted for images within the same observation.

The observed random variations and the limited amount of Phobos observations prevent us from performing any systematic correction to the whole VMC archive. The effect of this time shift can be neglected in many science analyses, which is particularly true for observations around apocenter, where due to the slow movement of the spacecraft, the fixed attitude during most observations, and the low VMC resolution, ~10 s are not enough to cause significant changes in the observed geometry. However, when using VMC data, especially images closer to pericenter, this shift should be taken into account, and subtracting ~10 seconds from the registered timestamp can be a good solution when inaccuracies in the geometry arise. Still, an uncertainty of ~4 seconds remains, and its effects need to be considered carefully in each case.

## 6. Summary and Perspectives

VMC was an engineering camera that has been recommissioned to be a science instrument. Within the limitations of the basic hardware that constitutes the instrument, its scientific performance is very good and the instrument is enabling novel science. This is in part because VMC provides observations over a wide FOV in the non-sun-synchronous elliptical orbit of MEX. The low data volume associated with the images allows the acquisition of sequences of images that can be used to investigate atmosphere dynamics on short-time scales and for measurements of wind.

Although the camera was not calibrated before launch, several campaigns to obtain in-flight dark currents, flat-fields and additional images allowed us to provide a basic photometric calibration (section 4), and a geometric characterization of the pointing and FOV (section 5). Although some hardware limitations remain, many of these can be overcome with operational strategies (section 3) and image processing techniques to obtain high level data products (section 4.3). Fig. 12 presents examples of the processed images that result from the work described in this paper.

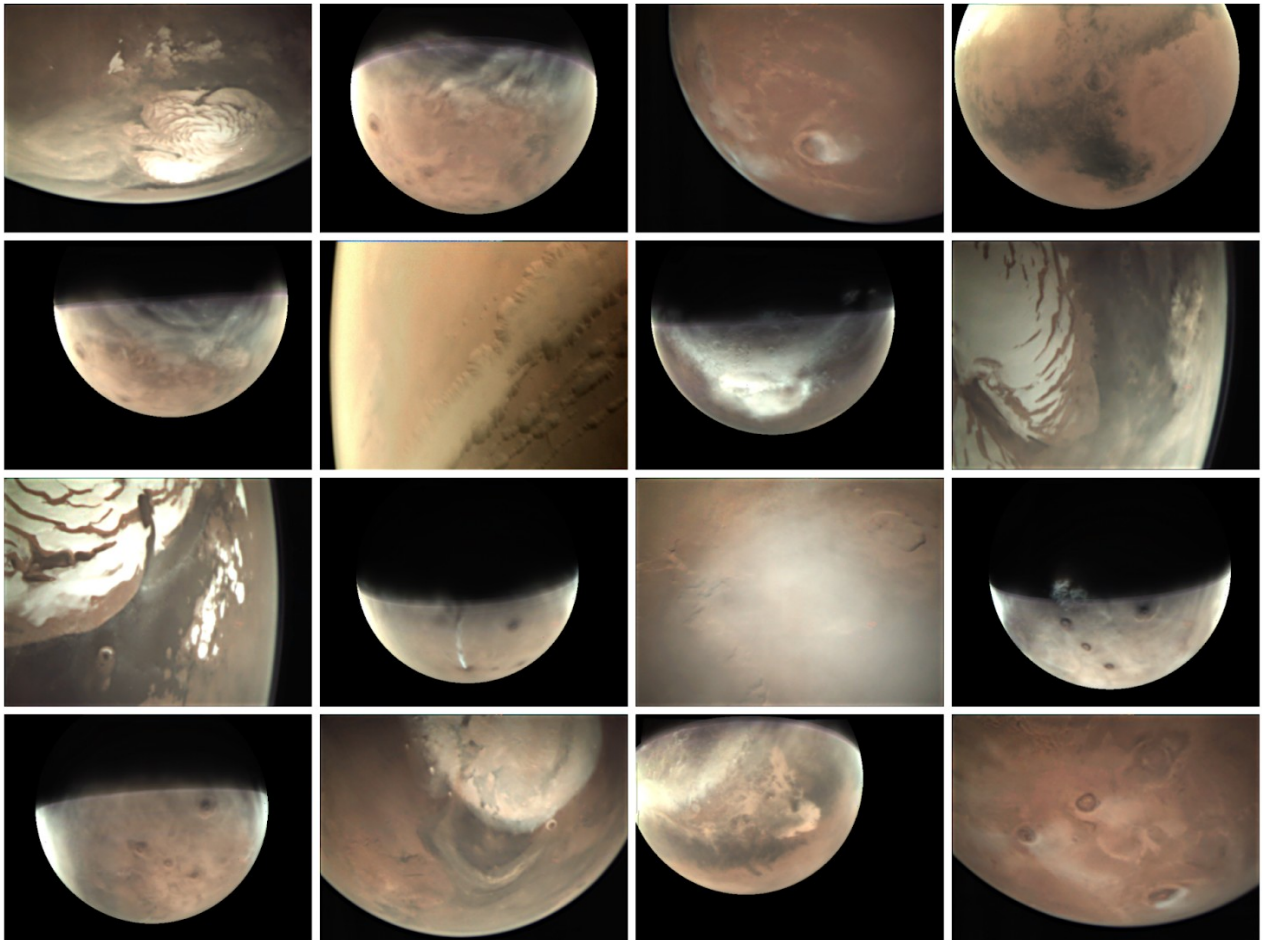

Fig. 12. A diverse collection of VMC images result from the work described in this paper. All these images are the result of calibration techniques (section 4.1) and HDR techniques (section 4.3), which require the accurate geometry determined in section 5. All of them are the result of extensive and coordinated work in operations (section 3).

Some imaging instruments currently in use on other missions around Mars share some capabilities with VMC. The IUVS (Imaging Ultraviolet Spectrograph; McClintock et al., 2015) instrument onboard MAVEN regularly obtains full disk images from apocenters. The EXI (Sharaf et al., 2020) camera onboard HOPE can obtain images of the full disk of Mars showing different local times and can also obtain images suitable for the analysis of winds. The HRSC (Jaumann et al., 2007) instrument onboard Mars Express has been obtaining push-broom scans from apocenters in recent years[3], enabling the monitoring of large areas at different local times, and there are ongoing experiments to take double push-broom scans suitable for cloud tracking. The recently retired MCC (Mars Color Camera; Arya et al., 2015) camera onboard MOM had a wide FOV and a high orbit that allowed the monitoring of wide areas at different local times. This instrument camera was also able to obtain videos suitable for wind studies from tracking clouds and dust patterns in the atmosphere.

Finding new science capabilities for auxiliary instruments not initially designed for science observations, like star trackers, is now taking place on different missions (Toldbo et al., 2022; Becker et al., 2020). In other cases, cameras originally included for outreach or citizen science purposes, like the Junocam instrument on the NASA Juno mission to Jupiter, have been used in several scientific applications (Hansen et al., 2017; Adriani et al., 2018). The inclusion of small cameras onboard planetary missions is now cheaper than it was when Mars Express was launched. Almost any small commercial camera now in the market is much better in performance and much lighter than VMC. Small cameras onboard planetary missions can provide interesting results both for science and outreach at a low cost. From a more general perspective, there are ongoing proposals to use full-disk imagers (Montabone et al., 2022) to improve the global monitoring of Mars, both in terms of spatial coverage and temporal resolution. Finally, wide FOV cameras could provide valuable observations of meteors in extraterrestrial atmospheres (Christou et al., 2012).

---

3   https://www.esa.int/Science_Exploration/Space_Science/Mars_Express/From_clouds_to_craters

## Acknowledgment

This work has been supported by Grant PID2019-109467GB-I00 funded by MCIN/AEI/10.13039/501100011033/ and by Grupos Gobierno Vasco IT-1742-22. Parts of this work were also funded by the Aula EspaZio Gela which is supported by a grant from the Diputación Foral de Bizkaia (BFA). JHB was supported by ESA Contract No. 4000118461/16/ES/JD, Scientific Support for Mars Express Visual Monitoring Camera and through the Faculty of the European Space Astronomy Centre (ESAC) - Funding reference ESAC-531. ACM also supported by grant PID2022-137579NB-I00 funded by MCIN/AEI/10.13039/501100011033 and by "ERDF A way of making Europe".

We acknowledge Bernhard Geiger for his experienced support in our discussions for the data calibration procedures. We acknowledge amateurs and space enthusiasts following and supporting VMC activities since its beginnings as an outreach instrument, especially members of the Unmanned Space Flight Forum in the early years.

## References

- Acton, C. H., Jr (1996). Ancillary data services of NASA's navigation and ancillary information facility. Planetary and Space Science, 44(1), 65– 70. doi: 10.1016/0032-0633(95)00107-7
- Acton, C., Bachman, N., Semenov, B., & Wright, E. (2018). A look towards the future in the handling of space science mission geometry. Planetary and Space Science, 150, 9–12. doi: 10.1016/j.pss.2017.02.013
- Adriani, A., Mura, A., Orton, G., Hansen, C., Altieri, F, Moriconi, M. L., … & Tosi, F. (2018). Clusters of cyclones encircling Jupiter's poles. Nature 555, 216-219, doi: 10.1038/nature25491
- Arya, A. S., Sarkar, S. S., Srinivas, A. R., Moorthi, S. M., Patel, V. D., Singh, R. B., … & Shah, D. (2015). Mars Colour Camera: the payload characterization/calibration and data analysis from Earth imaging phase. Current Science, 109.
- S. Barabash, R. Lundin, H. Andersson et al. (2004). ASPERA-3: Analyser of Space Plasmas and Energetic Ions for Mars Express. In: Mars Express: the scientific payload. (Vol. 1240, pp. 121–140).
- Becker, H. N., Alexander, J. W., Atreya, S.K., Bolton, S. J., Brennan, M. J., Brown, S. T., … & Steffes, P. G. (2020). Small lightning flashes from shallow electrical storms on Jupiter. Nature 584, 55–58. doi: 10.1038/s41586-020-2532-1
- Bellucci, G., Altieri, F., Bibring, J. P., Bonello, G., Langevin, Y., Gondet, B., & Poulet, F. (2006). OMEGA/Mars Express: Visual channel performances and data reduction techniques. Planetary and Space Science, 54(7), 675-684.

- J.-L. Bertaux, D. Fonteyn, O. Korablev et al. (2004). SPICAM: Studying the Global Structure and Composition of the Martian Atmosphere. In: Mars Express: the scientific payload. (Vol. 1240, pp. 95–120).
- Besse, S., Vallat, C., Barthelemy, M., Coia, D., Costa, M., De Marchi, G., … & Vallejo, F. (2018). ESA's Planetary Science Archive: Preserve and present reliable scientific data sets. Planetary and Space Science, 150, 131-140. https://doi.org/10.1016/j.pss.2017.07.013
- Bibring, J. P., Soufflot, A., Berthé, et al. (2004). OMEGA: Observatoire pour la Minéralogie, l'Eau, les Glaces et l'Activité. In: Mars Express: the scientific payload. (Vol. 1240, pp. 37–51).
- Cardesin-Moinelo, A., Godfrey, J., Grotheer, E. et al. Mars Express: 20 Years of Mission, Science Operations and Data Archiving. Space Sci Rev 220, 25 (2024). https://doi.org/10.1007/s11214-024-01059-0
- Chicarro, A., Martin, P., & Trautner, R. (2004). The Mars Express mission: an overview. In: Mars Express: the scientific payload. (Vol. 1240, pp. 3–16).
- Christou, A. A., et al. "Orbital observations of meteors in the Martian atmosphere using the SPOSH camera." Planetary and Space Science 60 (2012): 229-235. doi: 10.1016/j.pss.2011.09.002
- Jameux, D. (2003, January 17). VMC Flight User Manual. http://archives.esac.esa.int/psa/ftp/MARS-EXPRESS/VMC/MEX-M-VMC-3-RDR-EXT7-V1.0/DOCUMENT/MEX-ESA-VMC_FLIGHT_MANUAL.PDF
- ESA/UPV-EHU/DADPS. Mars Express Visual Monitoring Camera (VMC) Experiment to Archive Interface Control Document (EAICD). http://archives.esac.esa.int/psa/ftp/MARS-EXPRESS/VMC/MEX-M-VMC-2-EDR-EXT7-V1.0/DOCUMENT/EAICD_VMC.PDF
- ESA SPICE Service. Mars Express SPICE Kernel Dataset. Operational Dataset. DOI: 10.5270/esa-trn5vp1
- V. Formisano, D. Grassi, R. Orfei et al. (2004). PFS: the Planetary Fourier Spectrometer for Mars Express. In: Mars Express: the scientific payload. (Vol. 1240, pp. 71–94).
- Habinc, S., Karlsson, A., Wijmans, W., Jameux, D., Ogiers, W., & de Vos, L. (2000). In-flight results using visual monitoring cameras. European Space Agency, ESA SP-457, 2000., p.71
- Hansen, C.J., Caplinger, M.A., Ingersoll, A., Ravine, M.A., Jensen, E., Bolton, S., Orton, G. (2017). Junocam: Juno's Outreach Camera. *Space Science Reviews* 213, 475–506 (2017). doi: 10.1007/s11214-014-0079-x.
- Hernández-Bernal, J., Sánchez-Lavega, A., del Río-Gaztelurrutia, T., Hueso, R., Cardesín-Moinelo, A., Ravanis, E. M., … & Wood, S. (2019). The 2018 Martian global dust storm over the South Polar Region studied with MEx/VMC. Geophysical Research Letters, 46(17-18), 10330-10337. doi: 10.1029/2019GL084266
- Hernández-Bernal, J., Sánchez-Lavega, A., del Río-Gaztelurrutia, T., Ravanis, E., Cardesín-Moinelo, A., Connour, K., … & Hauber, E. (2021a). An Extremely Elongated Cloud over Arsia Mons Volcano on Mars: I. Life Cycle. Journal of Geophysical Research: Planets, 126(3), e2020JE006517. doi: 10.1029/2020JE006517
- Hernández-Bernal, J., Sánchez-Lavega, A., del Río-Gaztelurrutia, T., Hueso, R., Ravanis, E., Cardesín-Moinelo, A., … & Titov, D. (2021b). A Long-Term Study of Mars Mesospheric Clouds Seen at Twilight Based on Mars Express VMC Images. Geophysical Research Letters, 48(7), e2020GL092188. doi: 10.1029/2020GL092188
- Hernández-Bernal, J., Sánchez-Lavega, A., Del Río-Gaztelurrutia, T., Hueso, R., Cardesín-Moinelo, A., de la Parra, J. M. Y., … & Titov, D. (2021c). Looking for Meteors and Fireballs in the atmosphere of Mars from the Visual Monitoring Camera (VMC) on Mars Express (No. EPSC2021-515). Copernicus Meetings. doi: 10.5194/epsc2021-515
- Jaumann, R., Neukum, G., Behnke, T., Duxbury, T. C., Eichentopf, K., Flohrer, J., … & Hoffmann, H. (2007). The high-resolution stereo camera (HRSC) experiment on Mars Express: Instrument aspects and experiment conduct from interplanetary cruise through the nominal mission. Planetary and Space Science, 55(7-8), 928-952. doi: 10.1016/j.pss.2006.12.003
- McClintock, W. E., Schneider, N. M., Holsclaw, G. M., Clarke, J. T., Hoskins, A. C., Stewart, I., … & Deighan, J. (2015). The imaging ultraviolet spectrograph (IUVS) for the MAVEN mission. Space Science Reviews, 195(1-4), 75-124. doi: 10.1007/s11214-014-0098-7
- Merritt, D. R., Cardesin Moinelo, A., Marin Yaseli de la Parra, J., Breitfellner, M., Blake, R., Castillo Fraile, M., … & Titov, D. (2018). Mars Express Science Operations During Deep Eclipse: An Example of Adapting Science Operations On Aging Spacecraft. In 2018 SpaceOps Conference (p. 2354).

- Montabone, L., Heavens, N., Cardesín-Moinelo, A., Forget, F., Guzewich, S., Karatekin, Ö., … (2022). The case and approach for continuous, simultaneous, quasi-global weather monitoring on Mars. 44th COSPAR Scientific Assembly 2022.
- Neukum, G., Jaumann, R. and the HRSC Co-Investigator and Experiment Team (2004). HRSC: The high resolution stereo camera of Mars Express. In: Mars Express: the scientific payload. (Vol. 1240, pp. 17–36).
- Ormston, T., Denis, M., Peschke, S., & Schulster, J. (2008). From Monitoring Camera to Mars Webcam-Producing Outreach from Ops. In SpaceOps 2008 Conference (p. 3297). doi: 10.2514/6.2008-3297
- Ormston, T., et al. "An ordinary camera in an extraordinary location: Outreach with the Mars Webcam." Acta Astronautica 69.7-8 (2011): 703-713. doi: 10.1016/j.actaastro.2011.04.015
- M. Pätzold, F.M. Neubauer, L. Carone et al. (2004). MaRS: Mars Express Orbiter Radio Science. In: Mars Express: the scientific payload. (Vol. 1240, pp. 141–1164).
- G. Picardi, D. Biccari, R. Seu et al. (2004). MARSIS: Mars Advanced Radar for Subsurface and Ionosphere Sounding. In: Mars Express: the scientific payload. (Vol. 1240, pp. 51–70).
- D. Pullan, M.R. Sims, I.P. Wright et al. (2004). Beagle 2: the Exobiological Lander of Mars Express. In: Mars Express: the scientific payload. (Vol. 1240, pp. 165–206).
- Ravanis, E. M., Cardesín-Moinelo, A., Hernandez-Bernal, J., Almeida, M., Wood, S., Sánchez-Lavega, A., … & Martin, P. (2019). Mars Express Visual Monitoring Camera: New Operations and Data Processing for more Science. EPSC2019, 15-20.
- Ravanis, E., Hernández-Bernal, J., Cardesín-Moinelo, A., Sánchez-Lavega, A., del Río-Gaztelurrutia, T., Hueso, R., … & Martin, P. (2020, September). From engineering to science: Mars Express Visual Monitoring Camera's first science data release. In European Planetary Science Congress (pp. EPSC2020-437). doi: 10.5194/epsc2020-437
- del Río-Gaztelurrutia, T., Sánchez-Lavega, A., Hernández-Bernal, J., Angulo, A., Hueso, R., Cardesín-Moinelo, A., Martin, P., Wood, S., and Titov, D. (2021): Analysis of the evolution of Martian polar caps during Martian Years 34-35 from Mars Express Visual Monitoring Camera, EGU General Assembly 2021, online, 19–30 Apr 2021, EGU21-7787 doi: 10.5194/egusphere-egu21-7787
- Sánchez-Lavega, A., Rogers, J.H., Orton, G.S., Garcia-Melendo, E., Legarreta, J. Colas, F. …, & Wesley, A. (2017). A planetary-scale disturbance in the most intense Jovian atmospheric jet from JunoCam and ground-based observations. Geophysical Research Letters, 44, 4679-4686 (2017).
- Sánchez-Lavega, A., Chen-Chen, H., Ordonez-Etxeberria, I., Hueso, R., del Rio-Gaztelurrutia, T., Garro, A., … & Wood, S. (2018). Limb clouds and dust on Mars from images obtained by the Visual Monitoring Camera (VMC) onboard Mars Express. Icarus, 299, 194-205. doi: 10.1016/j.icarus.2017.07.026
- Sánchez-Lavega, A., Garro, A., del Río-Gaztelurrutia, T., Hueso, R., Ordoñez-Etxeberria, I., Chen Chen, H., … & Gondet, B. (2018). A seasonally recurrent annular cyclone in Mars northern latitudes and observations of a companion vortex. Journal of Geophysical Research: Planets, 123(11), 3020-3034. doi: 10.1029/2018JE005740
- Sánchez-Lavega, A., Erkoreka, A., Hernández-Bernal, J., del Río-Gaztelurrutia, T., García-Morales, J., Ordoñez-Etxeberría, I., … & Matz, K. D. (2022a). Cellular patterns and dry convection in textured dust storms at the edge of Mars North Polar Cap. Icarus, 387, 115183. doi: 10.1016/j.icarus.2022.115183
- Sanchez-Lavega, A., Larsen, E., Hernandez-Bernal, J., del Río-Gaztelurrutia, T., Ordoñez-Etxeberría, I., Ordorika, J., and Cardesín-Moinelo, A. (2022b): Atmospheric disturbances imaged on Mars during the simultaneous operations of four surface stations along 2021 and 2022, Europlanet Science Congress 2022, Granada, Spain, 18–23 Sep 2022, EPSC2022-228, 2022. doi: 10.5194/epsc2022-228
- Sánchez-Lavega, A., del Río-Gaztelurrutia, T., Spiga, A., Hernández-Bernal, J., Larsen, E., Tirsch, D., … & Machado, P. (2024). Dynamical phenomena in the Martian atmosphere through Mars Express imaging. Space Science Reviews, 220(1), 16.
- Scholten, F., Gwinner, K., Roatsch, T., Matz, K. D., Wählisch, M., Giese, B., … & Neukum, G. (2005). Mars Express HRSC data processing–Methods and operational aspects. Photogrammetric Engineering & Remote Sensing, 71(10), 1143-1152.
- Schulster, J., Gondet, B., Witasse, O., Denis, M., Griebel, H., & Ormston, T. (2011, October). Support to the CO2 Cloud Observations by Mars Express with the VMC Visual Monitoring Camera. In Proceedings of the EPSC-DPS Joint Meeting (Vol. 6).

- Sharaf, O., Amiri, S., Al Dhafri, S., Withnell, P., & Brain, D. (2020). Sending hope to Mars. Nature Astronomy, 4(7), 722-722. doi: 10.1038/s41550-020-1151-y
- Toldbo, C., Sushkova, J., Herceg, M., Denver, T., Benn, M., Jørgensen, P.S., ...& Strømme, A. (2022). Mapping High Energy Particles Using Augmented Star Trackers On-Board Swarm. Space Science Reviews, 218, 58. doi: 10.1007/s11214-022-00925-z

# Supplementary materials for

The Visual Monitoring Camera (VMC) on Mars Express: a new science instrument made from an old webcam orbiting Mars

The following supplementary materials are given below:

- Details on observations used to analyze the background noise (dark/bias).
- Details on observations used to create the flat field.

## Details on observations used to analyze the background noise (dark/bias)

This is the list of IDs of the images acquired for this analysis, described in section 4.1.1.

MEXVMC_1804330001
MEXVMC_1804330002
MEXVMC_1804330003
MEXVMC_1804330004
MEXVMC_1804330005
MEXVMC_1804330006
MEXVMC_1804330007
MEXVMC_1804330008
MEXVMC_1804370001
MEXVMC_1804370002
MEXVMC_1804370003
MEXVMC_1804370004
MEXVMC_1804370005
MEXVMC_1804370006
MEXVMC_1804370007
MEXVMC_1804370008

## Details on observations used to create the flat field

This is the list of IDs of 120 images used to generate the flat field, with the procedure described in section 4.1.2.

MEXVMC_1000070002
MEXVMC_1000070006
MEXVMC_1000070010
MEXVMC_1000070014
MEXVMC_1000070018
MEXVMC_1000070022
MEXVMC_1000070026
MEXVMC_1000070030
MEXVMC_1701310003
MEXVMC_1701310004
MEXVMC_1701310006
MEXVMC_1701310007
MEXVMC_1701310008
MEXVMC_1701310009
MEXVMC_1701310010
MEXVMC_1701310011
MEXVMC_1701310012
MEXVMC_1701360001
MEXVMC_1701360003
MEXVMC_1701360004
MEXVMC_1701360005
MEXVMC_1701360006
MEXVMC_1701360007
MEXVMC_1701360008
MEXVMC_1701360009
MEXVMC_1701360010
MEXVMC_1701360011
MEXVMC_1701360012
MEXVMC_1701360013
MEXVMC_1701360014
MEXVMC_1701360015
MEXVMC_1701360016
MEXVMC_1701360017
MEXVMC_1701360018
MEXVMC_1701360019
MEXVMC_1701360020
MEXVMC_1701360021
MEXVMC_1701360022
MEXVMC_1701400003
MEXVMC_1701400004
MEXVMC_1701400006
MEXVMC_1701400007

MEXVMC_1701400009
MEXVMC_1701400010
MEXVMC_1701400011
MEXVMC_1701400012
MEXVMC_1701400013
MEXVMC_1701400014
MEXVMC_1701400015
MEXVMC_1701400016
MEXVMC_1701400017
MEXVMC_1701400018
MEXVMC_1701400019
MEXVMC_1701400020
MEXVMC_1701400021
MEXVMC_1701400022
MEXVMC_1701420004
MEXVMC_1701420006
MEXVMC_1701420007
MEXVMC_1701420009
MEXVMC_1701420010
MEXVMC_1701420011
MEXVMC_1701420012
MEXVMC_1701420013
MEXVMC_1701420014
MEXVMC_1701420015
MEXVMC_1701420016
MEXVMC_1701420017
MEXVMC_1701420018
MEXVMC_1701420019
MEXVMC_1701420020
MEXVMC_1701420021
MEXVMC_1701420022
MEXVMC_1701420023
MEXVMC_1701420024
MEXVMC_1701420025
MEXVMC_1701420026
MEXVMC_1701420027
MEXVMC_1701420028
MEXVMC_1701420029
MEXVMC_1701420030
MEXVMC_1701420031
MEXVMC_1701420032
MEXVMC_1701420033
MEXVMC_1701420034
MEXVMC_1701420035
MEXVMC_1701590007

MEXVMC_1701590009
MEXVMC_1701590010
MEXVMC_1701610007
MEXVMC_1701610009
MEXVMC_1701610010
MEXVMC_1701660007
MEXVMC_1701660009
MEXVMC_1701660010
MEXVMC_1701680010
MEXVMC_1701690007
MEXVMC_1701690009
MEXVMC_1701690010
MEXVMC_1701820010
MEXVMC_1801390001
MEXVMC_1801390002
MEXVMC_1801390003
MEXVMC_1801390004
MEXVMC_1801390005
MEXVMC_1801390006
MEXVMC_1801390007
MEXVMC_1801390008
MEXVMC_1801390009
MEXVMC_1801390010
MEXVMC_1801440001
MEXVMC_1801440002
MEXVMC_1801440003
MEXVMC_1801440004
MEXVMC_1801440005
MEXVMC_1801440006
MEXVMC_1801440007
MEXVMC_1801440008
MEXVMC_1801440009
MEXVMC_1801440010